\documentstyle[epsfig,aps,preprint,prb,tighten]{revtex}

\begin{document}

\title{Local lattice disorder in the geometrically-frustrated spin glass pyrochlore
Y$_2$Mo$_2$O$_7$}

\author{C. H. Booth,$^{1,2}$ J. S. Gardner,$^2$ G. H. Kwei,$^2$ 
R. H. Heffner,$^2$ F. Bridges,$^3$ and M. A. Subramanian$^4$}

\address{$^1$ Chemical Sciences Division, Lawrence Berkeley National Laboratory,
Berkeley, California 94720}
\address{$^2$ Los Alamos National Laboratories, Los Alamos, New Mexico, 87545}
\address{$^3$ Department of Physics, University of California, Santa Cruz, 
California, 95064} 
\address{$^4$ Dupont Central Research and Development, Experimental Station, 
Wilmington, Delaware 19880}

\date{submitted to Phys. Rev. 10/22/99}

\maketitle

\begin{abstract}

The geometrically-frustrated spin glass Y$_2$Mo$_2$O$_7$ has been 
considered widely to be crystallographically ordered with a
unique nearest neighbor magnetic exchange interaction, $J$.
To test this assertion, we present x-ray-absorption fine-structure 
results for the Mo and Y
$K$ edges as a function of temperature and 
compare them to results from a well-ordered pyrochlore, Tl$_2$Mn$_2$O$_7$.
We find that the Mo-Mo pair distances are significantly disordered
at approximately right angles to the Y-Mo pairs.
These results strongly suggest that lattice disorder nucleates the spin-glass 
phase in this material.

\end{abstract}




Materials with magnetic ions that reside on the corners of triangles 
or tetrahedra and have a nearest-neighbor antiferromagnetic interaction 
$J$ potentially display phenomena known broadly as 
{\it geometrical frustration}.\cite{FrustratedReviews}
Such 
symmetries preclude a  simple two-sublattice N\'eel
ordering, as it is not possible to satisfy all near-neighbor
antiferromagnetic interactions.  Materials that exhibit frustration 
can still undergo a magnetic transition whereby spins freeze into a lattice 
without long-range order, that is, they form a spin glass (SG).
Theoretical calculations have long asserted that a SG transition is 
only possible when both frustration (potentially 
induced by lattice disorder) and a distribution of effective $J$'s exists.
\cite{NeedDisorder}
Many earlier experimental studies bear this assertion 
out.\cite{Fisher91}  Recently, there has been great interest in some 
SG materials that apparently contradict this 
assertion.\cite{FrustratedReviews}   Such systems include the 
Gd$_3$Ga$_5$O$_{12}$ (GGG) \cite{Schiffer95b} and some
pyrochlores such as Y$_2$Mo$_2$O$_7$.
\cite{Reimers88,Raju92,Gingras97,Dunsiger96,Gardner99,Walton99}
In the case of GGG, some
off-stoichiometry is known to exist, but is likely too little to explain the SG 
transition using conventional methods.  Y$_2$Mo$_2$O$_7$, on the other 
hand, is apparently pure, stoichiometric, and crystallographically well 
ordered.\cite{Reimers88}  Consequently, it has been studied with
a wide range of techniques.  This system displays strong indications of a
SG transition at $T_{\rm g}$=22.5 K, including irreversible behavior in 
the bulk magnetic susceptibility \cite{Raju92,Gingras97} and a rapid slowing 
down and freezing of the magnetic moments with no long-range order down to 
$\sim 0.1$~$T_{\rm g}$ from neutron scattering and muon spin 
rotation/relaxation ($\mu$SR) data.\cite{Dunsiger96,Gardner99} However,
some magnetic properties are anomalous when compared to conventional SG 
materials.\cite{Dunsiger96,Walton99}  Such properties seem to point the way 
towards some form of new mechanism for SG formation.

Frustration arises in the Mo sublattice of Y$_2$Mo$_2$O$_7$ because the Mo 
atoms occupy a
corner-shared tetrahedral network, making this system a three 
dimensional analog of the kagom\'{e} lattice of a corner-shared triangular 
network.  
Although not generally recognized, Y$_2$Mo$_2$O$_7$ has a substantially enhanced
displacement parameter for the transition-metal site of 
$<u^2> \approx$0.0145 \AA$^2$ at room temperature \cite{Reimers88}
(compare to 0.0075 \AA$^2$ 
for the Mn site in the well-ordered pyrochlore 
Tl$_2$Mn$_2$O$_7$\cite{Subramanian96}).  Such a large displacement parameter 
could either be due to an unusually low Debye temperature (i.e. large dynamic 
displacements due to thermal phonons) for the Mo site, or
to static (positional) displacements that are not described well by the 
standard Rietveld 
refinement.  These facts
lead us to search for pair-distance disorder in 
Y$_2$Mo$_2$O$_7$ using the x-ray-absorption fine-structure (XAFS) 
technique as a local probe of the Y and Mo environments.  
Using information about both 
these environments and comparing to data on 
Tl$_2$Mn$_2$O$_7$, we show below that the Mo tetrahedra are, in fact, 
disordered from their average, ideal structures.  Furthermore, the magnitude 
of this disorder is roughly consistent with the SG transition 
temperature $T_{\rm g}$ and mean-field theory (MFT).\cite{Sherrington75a}


The polycrystalline sample is the same sample used in Ref. 
\onlinecite{Gardner99},
and was prepared and characterized in a similar manner 
to that reported previously.\cite{Reimers88,Raju92,Gingras97,Dunsiger96}
In particular, field-cooled and zero field-cooled magnetization measurements
were performed which displayed the signature irreversible behavior and a 
$T_{\rm g}=22.5$ K, indistinguishable from that reported 
previously.\cite{Gingras97}
A neutron powder diffraction study indicates the deviation from nominal oxygen 
stoichiometry is less than 1\%,\cite{Gardner99} and is consistent with
previous studies in all respects, including the $<u^2>$ 
parameters.\cite{Reimers88}  Before collecting x-ray
absorption data, the sample was ground under acetone, sifted
through a 30 $\mu$m sieve, and brushed onto scotch tape.
Strips of tape were stacked such that the Mo $K$ (20 keV) and Y $K$ (17 keV) 
absorption steps were about one absorption length.  The samples were loaded 
into a LHe flow cryostat, and 
absorption spectra were collected between 15-300~K on BL~4-1 at the Stanford 
Synchrotron Radiation Laboratory (SSRL) using Si(220) monochromator crystals.

The XAFS technique involves measuring and analyzing the absorption spectrum 
just above an absorption edge.
Oscillations occur when the outgoing 
photoelectron (p.e.) is partially backscattered by neighboring atoms, 
interfering with the outgoing component.  The interference depends on 
the p.e. wave vector $k$, and
on the pair-distance distribution of the various atomic shells.  An 
oscillatory function, $\chi(k)$, is extracted and the Fourier
transform (FT) of $\chi(k)$ gives peaks in $r$-space that correspond to 
various coordination shells.  FT's are not exactly radial-distribution 
functions, however, and fits need to 
be performed to extract numerical structural information.  These fits provide 
the pair distance $R$ from the absorbing atomic species
to a given shell around that atom, the pair-distance distribution 
variance 
$\sigma^2$ and, in some cases, the skewness of the distribution,
which we report in terms of cumulants.\cite{Eisenberger79}

All data reduction 
and fits in this paper utilize the standard procedures in 
Ref. \onlinecite{Hayes82,Li95b,Bridges95b}.  
In particular, we fit the FT of $k^3\chi(k)$
to backscattering and phase functions calculated by
FEFF7.\cite{FEFF6}  The following results focus on single-scattering paths, 
although multiple scattering is included in the analysis. 
Each backscattering species was allowed to
have a different p.e. threshold energy.
Errors reported for the XAFS fit parameters $R$ and $\sigma^2$ are based on 
the reproducibility of the fits
for each of the three scans taken at each temperature.  Absolute errors for
these parameters are typically much larger,\cite{Li95b} about 10\% for 
nearest neighbors and roughly 20\% for further neighbors in $\sigma^2$, and
about 0.01 \AA~for nearest neighbors and 0.02 \AA{ } for 
further neighbors $R$.  

\begin{figure}
\epsfig{file=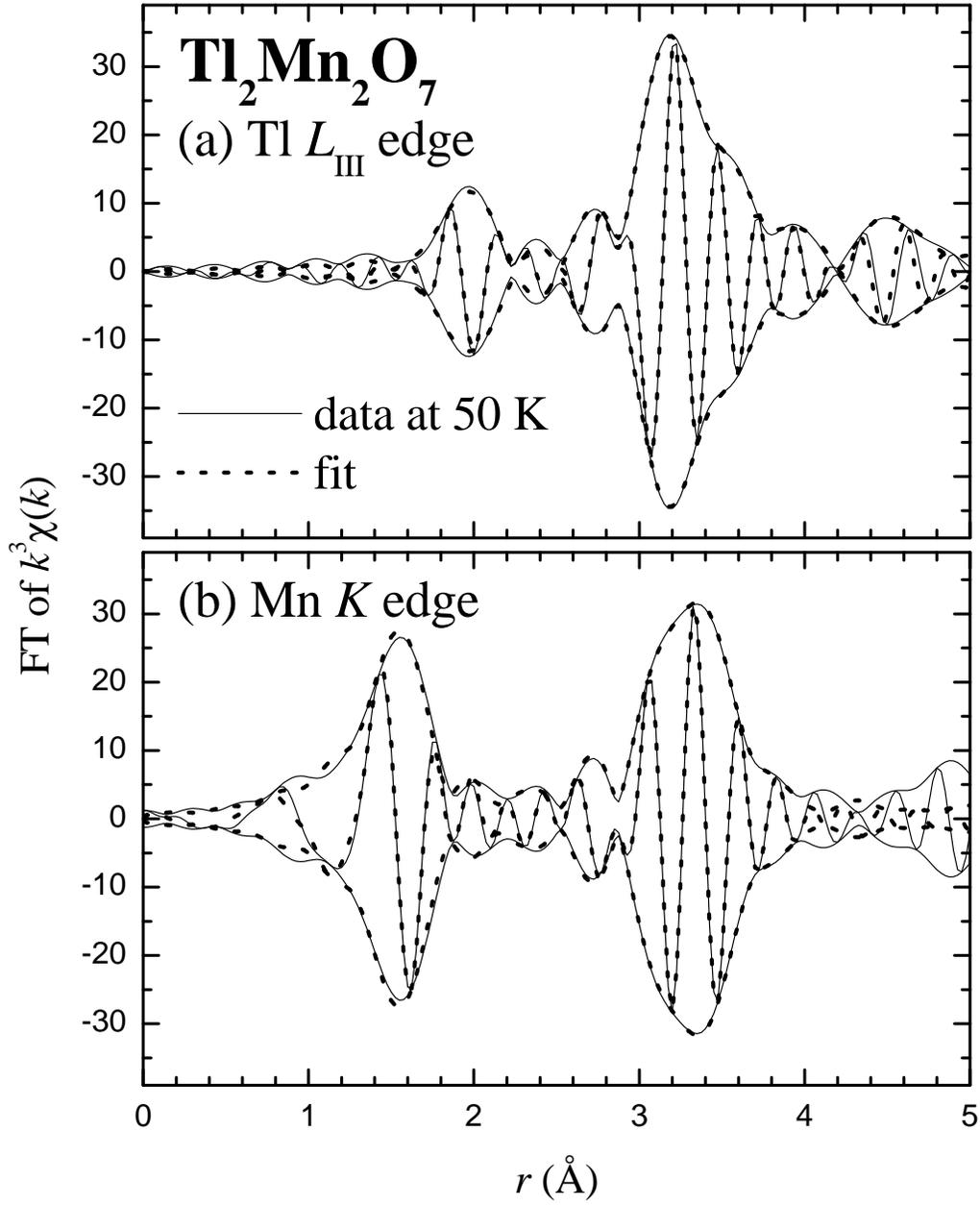, width=6.0in}
\caption{XAFS data and fits on Tl$_2$Mn$_2$O$_7$.
The outer envelope
shows $\pm$ the amplitude (or modulus) and the oscillating inner curve
is the real part of each complex transform.  The $r$-axis
includes different phase shifts for each coordination shell, so that the $r$
of a given peak does not equal the pair distance $R$.  Transforms are from 
3.0 and
15.0 \AA$^{-1}$, and Gaussian narrowed by 0.3~\AA$^{-1}$.
Fits are from 1.0 and 4.0 \AA.}
\label{TMO_fig}
\end{figure}

To demonstrate that an ordered pyrochlore can accurately be measured using 
XAFS, we chose to measure the ``colossal'' magnetoresistance material 
Tl$_2$Mn$_2$O$_7$.  These data were partially analyzed earlier.\cite{Kwei97} 
An example of the XAFS data at $T =$~50~K is presented in Fig. \ref{TMO_fig}.
The nominal structure is the same as  Y$_2$Mo$_2$O$_7$ with Tl and Mn 
replacing Y and Mo, respectively.  Rietveld refinements\cite{Subramanian96} 
show that each Tl site has six O(1) neighbors at 2.43~\AA~and two O(2)'s at 
2.15~\AA.  The XAFS from these oxygens interfere with each other, reducing 
their combined amplitude and producing the peak centered 
at about 2~\AA{ }in the FT data.  The oxygen coordination for Mn is similar 
with six O(1)'s at 1.90 \AA, but without any further nearest-neighbor 
oxygens, and therefore the XAFS has a much stronger peak in the FT at 
$\sim 1.6$ \AA.  The next-nearest-neighbor coordination for the Tl and Mn sites 
are the same, with six Mn and six Tl neighbors at 3.49 \AA~(a ``Tl/Mn" peak at 
$\sim 3.3$ \AA~in Fig. \ref{TMO_fig}).  These occupancies are assumed in the 
fits.

Table \ref{TMO_table} shows the fit results from the Tl $L_{III}$ and Mn
$K$ edge XAFS of Tl$_2$Mn$_2$O$_7$.  Data were collected from 50-300 K, 
and no abnormal temperature dependence in either the pair distances or 
distribution variance was observed.  The pair-distance distribution variance 
$\sigma^2$ for each pair of atoms is described well by a correlated-Debye 
model \cite{Crozier88} (i.e. phonons) with no appreciable static disorder.  The 
correlated-Debye model has been shown to describe the temperature dependence 
of $\sigma^2$ with no additional temperature independent offset (``static'') 
component for the well-ordered oxide CaMnO$_3$. 
If an offset is necessary, it is quantitatively equal to extra lattice
disorder as in La$_{1-x}$Ca$_x$MnO$_3$.\cite{Booth98a}  In any case, rather 
than show the temperature dependence of $\sigma^2$ for each shell, we merely 
list the correlated-Debye temperature $\Theta_{cD}$ and 
offset $\sigma_{\rm stat}^2$ in Table \ref{TMO_table}.

\begin{figure}
\epsfig{file=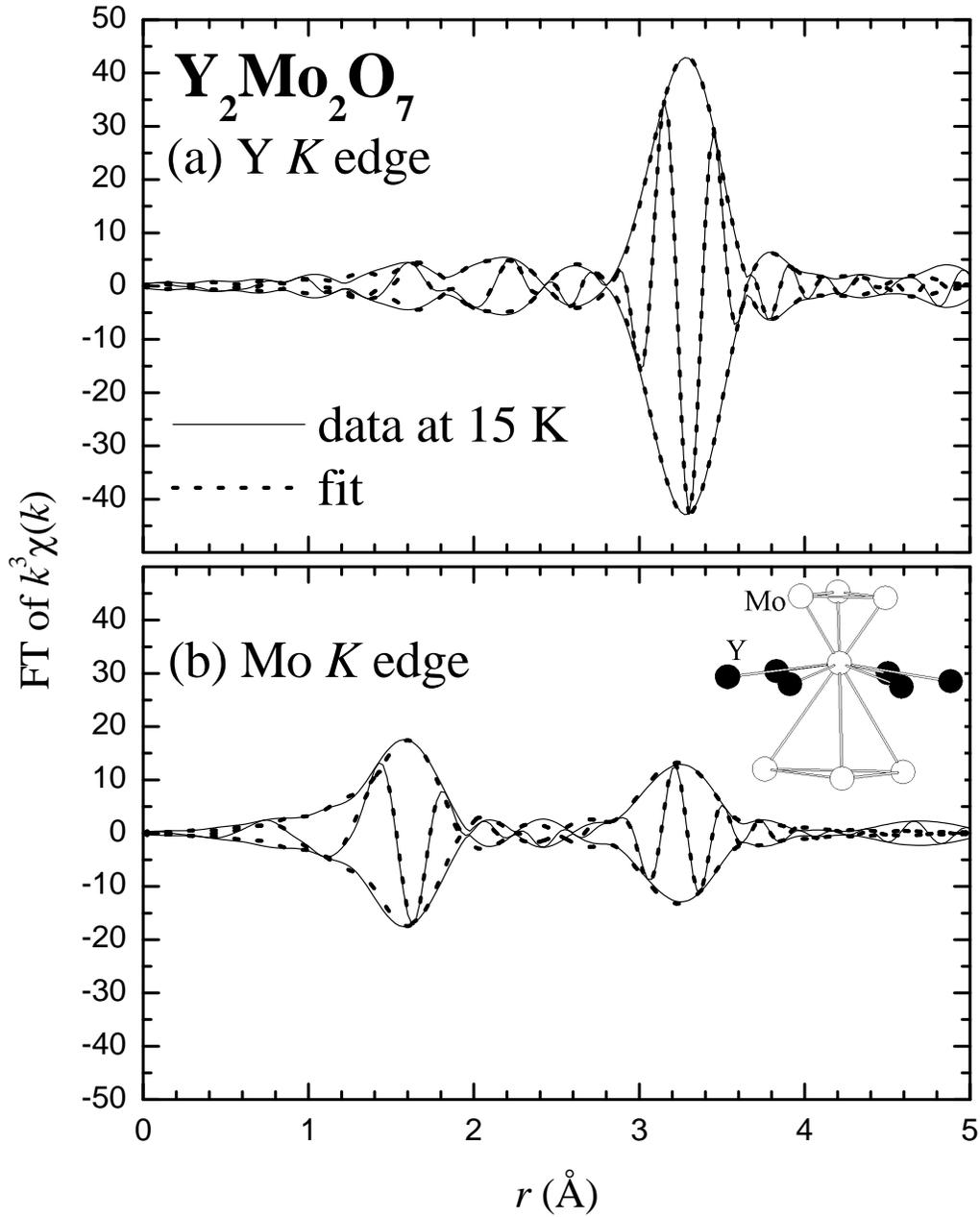, width=6.0in}
\caption{XAFS data and fits on Y$_2$Mo$_2$O$_7$.  Fit and transform ranges
are as in Fig. \protect\ref{TMO_fig}.  The inset shows one possible 
(exaggerated) distortion
of the Mo tetrahedra that might explain the difference in the peak amplitude at 
3.3 \AA{ } between the two edges.  Oxygens are omitted for clarity.  
}
\label{YMO_fig}
\end{figure}

Figure \ref{YMO_fig} shows the data and fits for Y$_2$Mo$_2$O$_7$ at 
$T=15$~K.
The raw data already strongly indicate some unexpected disorder in the Y/Mo 
peak.  Since these environments are nominally the same, this peak should have 
nearly the same amplitude for both the Mo and Y edge data.  However, as is 
seen in Fig. \ref {YMO_fig}, the Mo-Mo/Y peak is only about 25\% the size of 
the Y-Mo/Y peak.
Table \ref{YMO_table} summarizes these fits for $T=$~15 K.
Like the fits to the Tl$_2$Mn$_2$O$_7$ 
data, we find no abnormal atom-pair distances compared to the 
diffraction results.  However, we do measure a large temperature-independent 
contribution to 
$\sigma^2$ in several atom pairs, most significantly in the Mo-Mo pairs.  
Several atom pairs show a
$\sigma_{\rm stat}^2$ contribution that is significantly larger than the 
estimated errors.  In particular, the Mo-Mo peak is almost completely washed 
out by its estimated static variance of $\sim 0.03$~\AA$^2$, while the Y-O(1) and 
Y-Mo/Y pairs also display a modest static component to their variance.  
Since the Y-Mo pairs also include Mo atoms,
we conclude that the primary direction of the large disorder in the Mo-Mo
pairs is roughly 
perpendicular to the Y-Mo pairs and parallel to the Mo-Mo pairs.  Notice that,
in spite of the large measured disorder, the diffraction results are in rough 
agreement with these XAFS results:  If one assumes the Mo displacements are uncorrelated with each other, one expects the XAFS measured variance in the Mo-Mo 
pairs to be roughly $2<u^2> \approx 0.029 $ \AA$^2$, in agreement with the 
measured value of 0.026(5) \AA$^2$.

The particularly large magnitude of the Mo-Mo variance $\sigma^2$ led us to 
attempt to fit this pair assuming some form of anharmonic distribution.  
Including a split peak in the fit did not
produce satisfactory results.  Some improvement to the fit was made by 
including a significant skewness in the Mo-Mo distribution, with a 
third-cumulant $C_3 = -0.080\pm0.005$ \AA,\cite{Eisenberger79} which also 
decreases the fitted $\sigma^2_{15 \rm K}$ to about 0.01 \AA$^2$.  Notice 
that the total disorder is approximately unchanged in this fit.  However, 
the overall improvement to the fit was not very significant.
With any description, however, the conclusion is that the
Mo-Mo pairs cannot be completely described by a 
single harmonic potential.

Table \ref{YMO_table} shows that the Mo-O(1) octahedron remains a tightly bound
unit, with a high $\Theta_{\rm cD}$ and little static disorder.  
If the Mo disorder carries the O(1) octahedron with it, then the Y-O(1) 
pairs must also be disordered.  Furthermore, 
the Y-Mo pair should show some disorder if the Mo-Mo disorder
is so large.  Indeed, for both these pairs we do measure
some excess disorder in $\sigma_{\rm stat}^2$.  

Although the disorder in the Mo-Mo pairs is obvious in the XAFS data, at
this time we can only speculate why such disorder exists in Y$_2$Mo$_2$O$_7$ 
and not in Tl$_2$Mn$_2$O$_7$.  One possibility is that a very small 
($\lesssim 1\%$) number of oxygen vacancies create enough Mo$^{3+}$ ions to 
precipitate the measured disorder.  This situation is very unlikely because 
we then expect equally strong disorder in the MoO$_6$ octahedra,
which we do not observe.  Similarly, any significant Y/Mo site interchange seems
unlikely due to the observed order in the Y environment and the MoO$_6$ 
octahedra, as well as ionic size arguments.  Another more intriguing 
possibility is that the Mn 
tetrahedra in Tl$_2$Mn$_2$O$_7$ are ordered because of the metallic Tl 6$s$ 
band and its interaction with the Mn 3$d$ band.  This mechanism is not present 
in Y$_2$Mo$_2$O$_7$, so the more localized Mo 4$d$ band may allow the
disorder to develop.  Comparisons to data on the SG Tb$_2$Mo$_2$O$_7$
and the ferromagnet Y$_2$Mn$_2$O$_7$ may help clarify this issue.


The above measurements show that the primary
disorder involves only the Mo atoms in a direction that is roughly parallel 
to local Mo-Mo pairs and perpendicular to Mo-Y pairs.  One possible distortion
is shown in the inset to Fig. \ref{YMO_fig}(b) by displacing the Mo atom towards
or away from the 
tetrahedron body center.  The magnitude of this distortion may vary throughout
the solid, creating a distribution of Mo-Mo pair distances and not
severely altering the Mo-Y pairs.  
Displacing the Mo towards a tetrahedron 
corner will similarly distort the Mo-Mo pairs while having little effect on the 
Mo-Y environment.

The Mo disorder has a very important implication for understanding 
the SG behavior of Y$_2$Mo$_2$O$_7$: the
fundamental assertion that a SG transition at finite 
temperature should not occur without both frustration and disorder 
in $J$ now appears to hold for 
Y$_2$Mo$_2$O$_7$.  It is, however, important to ask whether the measured
disorder is enough to explain the magnitude of $T_{\rm g}$.  We first 
define the disorder in $J$ as 
$\sigma_J/J_0$ with a mean magnetic interaction $J_0$ and
a variation in $J$ of $\sigma_J^2$.  To estimate the effect of pair-distance 
disorder, we use the simple Heitler-London result that $J \propto 1/R$ with a 
variation $\sigma_R^2$ in the pair-distance distribution and a mean distance 
$R_0$.  This approximation gives $\sigma_J/J_0 \approx \sigma_R / R_0 
\approx 5\%$ for this XAFS measurement.  We then compare this value of the
disorder due to bond length variations to 
the disorder in $J$ estimated from the quantum-mechanical MFT 
with Ising spins of Sherrington and Southern (SS).\cite{Sherrington75a}
Using Eq. 12 in SS with a coordination $z=6$, $T_{\rm g} = 20 K$ and $S=2$ 
for a Mo$^{4+}$ ion, we obtain $\sigma_J=$~0.39 meV.  Similarly, using Eq. 17 
in SS together with the measured Curie-Weiss temperature $\Theta_{\rm CW}$ 
of 200 K,\cite{Gingras97} we obtain $J_0=$~1.74 meV.  Therefore, according 
to SS, we expect to find the same ratio $\sigma_J/J_0$ to be $\approx$ 20-25\%. 

When comparing the expected disorder of 20-25\% in the SS theory to that 
approximated from the pair-distance disorder of $\sim$5\%, it is important to 
remember that the SS theory uses many approximations, including Ising spins, 
nearest-neighbor-only interactions and MFT.  With these caveats, the fact that 
SS theory 
predicts the same order of magnitude of disorder in $J$ as estimated from the 
lattice disorder 
indicates strongly that lattice disorder plays a major role in the 
SG transition.  
On the other hand, the measured disorder is still a factor of 5 less than
expected from the SS arguments, and so it is very unclear whether 
Y$_2$Mo$_2$O$_7$ belongs in the class of low-disorder spin glasses such as GGG
or not.
Therefore it is very important that both Monte 
Carlo calculations be performed using the measured disorder and 
that other probes be used to clarify the exact nature of the disorder.  
Pair-distribution-function analysis of neutron or x-ray diffraction data might 
be very helpful in this regard once the general parameters of the disorder from 
the XAFS measurements are included.\cite{Billinge_priv}  Such Monte Carlo 
calculations should clarify if pair-distance disorder (as opposed to 
substitutional disorder) can explain any of the unconventional properties in
Y$_2$Mo$_2$O$_7$.\cite{Dunsiger96,Walton99}


In summary, we have performed temperature dependent XAFS experiments at both 
the Y and Mo $K$ edges on a well characterized sample of Y$_2$Mo$_2$O$_7$.  
These measurements indicate a relatively large amount of pair-distance disorder 
(between 
0.1 and 0.2 \AA) of the Mo atoms exists, roughly perpendicular to the Y-Mo 
pairs and possibly in the 
direction of the Mo tetrahedron face centers.  In addition, the oxygen 
octahedron surrounding the Mo site is roughly preserved, while the octahedron
surrounding the Y site is distorted.  These measurements indicate that 
lattice disorder does exist in these materials in the form of 
pair-distance disorder, and therefore the theoretical requirement that disorder 
in $J$ exist for a SG transition to 
occur appears to be met.  The degree of disorder is still low when 
compared to that expected from MFT, underlining the need for a new approach 
towards understanding low-disorder spin glasses both by better characterization 
of existing disorder and from theoretical descriptions that include these
characterizations.

We thank S. J. L. Billinge and J. A. Mydosh for useful conversations. 
This work
was partially supported by the Office of 
Basic Energy Sciences (OBES), Chemical Science Division of the 
Department of Energy (DOE), Contract no. DE-AC03-76SF00098.
Work at Los Alamos National Laboratory was conducted under the 
auspices of the DOE.  X-ray absorption data 
were collected at SSRL, which is operated by the DOE/OBES.


\begin{table}
\caption{Fit results for Tl and Mn edge data on Tl$_2$Mn$_2$O$_7$.  
$S_0^2$=1.0 for the Tl edge fits and 0.92 for the Mn edge fits.  
Results for $\sigma^2$ and $R$ are at $T$=50 K.  See text for a discussion of 
errors on $R$ and $\sigma^2$.
Error estimates for $\sigma_{\rm stat}^2$ and 
$\Theta_{\rm cD}$ are based on the covariance matrix from the least-squares fit
to $\sigma^2(T)$.
Superscripts * and \# indicate values constrained to be the same in the fits.}

\begin{tabular}{lcccc}
atom pair & $\sigma^2$ (\AA$^2$) & $R$ (\AA) & $\Theta_{\rm cD}$ (K) & $\sigma_{\rm stat}^2$ (\AA$^2$) \\
\tableline
Mn-O(1) & 0.0025(1) & 1.902(2) & 800(70) & 0.0001(3) \\
Mn-Mn   & 0.0037(1) & 3.492(2)$^*$ & 600(60) & 0.0014(4) \\
Mn-Tl   & 0.0026(1) & 3.492(2)$^*$ & 275(30) & 0.0001(1) \\
\\
Tl-O(1) & 0.0060(1) & 2.431(1) & 480(30) & 0.0008(6) \\
Tl-O(2) & 0.0011(2) & 2.146(1) & 850(90) &-0.0001(3) \\
Tl-Tl   & 0.0023(1) & 3.491(1)$^\#$ & 345(10) &-0.0002(2) \\
Tl-Mn   & 0.0024(1) & 3.491(1)$^\#$ & 230(20) & 0.0002(5) 

\end{tabular}
\label{TMO_table}
\end{table}

\begin{table}
\caption{Fit results for Y and Mo edge data on Y$_2$Mo$_2$O$_7$.  
$S_0^2$=1.0 for the Y edge and 0.83 for the Mo edge fits.  
Results for $\sigma^2$ and $R$ are at $T=$~15 K.  Note that $\sigma^2$'s 
and $R$'s
for Y-Y and Y-Mo were constrained together.  Also note that
$\Theta_{\rm cD}$ was held fixed at 260 K to estimate
$\sigma_{\rm stat}^2$, but that no resolvable temperature dependence
was observed.
}
\begin{tabular}{lcccc}
atom pair & $\sigma^2$ (\AA$^2$) & $R$ (\AA) & $\Theta_{\rm cD}$ (K) & $\sigma_{\rm stat}^2$ (\AA$^2$) \\
\tableline
Mo-O(1) & 0.0045(3) & 2.030(2) & 950(40) & 0.0024(1) \\
Mo-Mo   & 0.035(1) & 3.628(2)$^*$ & 260({\it fixed}) & 0.026(5) \\
Mo-Y    & 0.0055(1) & 3.628(2)$^*$ & 356(20) & 0.0033(4) \\
\\
Y-O(1) & 0.0092(3) & 2.441(1) & 525(12) & 0.0046(3) \\
Y-O(2) & 0.0036(1) & 2.225(1) & 860(60) & 0.0009(3) \\
Y-Y/Mo    & 0.0042(1) & 3.609(1) & 381(1) & 0.0020(2) 

\end{tabular}
\label{YMO_table}
\end{table}

\end{document}